\documentclass[smallextended]{svjour3} 
\usepackage[english]{babel}
\usepackage[utf8]{inputenc}
\usepackage{pgf,tikz,pgfplots}
\pgfplotsset{compat=1.17}
\usetikzlibrary{arrows}
\usepackage{amssymb}
\usepackage{bm}
\usepackage{csquotes}
\expandafter\let\csname equation*\endcsname\relax
\expandafter\let\csname endequation*\endcsname\relax
\usepackage{amsmath}
\usepackage{amsfonts}
\usepackage{amstext}
\usepackage{physics}
\usepackage{color}
\usepackage{hyperref}
\usepackage{enumerate}
\usepackage{graphicx}
\usepackage{float}
\usepackage{setspace} 
\usepackage{lipsum}
\usepackage{dcolumn}
\usepackage{caption}
\usepackage{subcaption}
\usepackage{epsfig}
\usepackage{epsf}
\usepackage{epstopdf}

\usepackage{enumitem}
\usepackage{xcolor}
\usepackage{tikz}
\usepackage[margin=1in]{geometry}
\usepackage{pgfplots}
\usepackage{esint}

\usepackage{scalerel}
\usepackage{stackengine}
\usepackage{braket}
\usepackage{MnSymbol}
\usepackage{titlesec}
\usepackage{verbatim}
\usepackage{tikz}

\DeclareSymbolFont{usualmathcal}{OMS}{cmsy}{m}{n}
\DeclareSymbolFontAlphabet{\mathcal}{usualmathcal}

\newcommand{\be}{\begin{equation}}
\newcommand{\ee}{\end{equation}}
\newcommand{\bea}{\begin{eqnarray}}
\newcommand{\eea}{\end{eqnarray}}

\usepackage{color}
\definecolor{dgreen}{rgb}{0,0.7,0}

\begin{document}
\title{Thermalization and hydrodynamics in an interacting integrable system: the case of hard rods}

\author{Sahil Kumar Singh\textsuperscript{1}, Abhishek Dhar\textsuperscript{1}, Herbert Spohn\textsuperscript{2,3} and Anupam Kundu\textsuperscript{1}}

\institute{${}^{1}$ International Centre for Theoretical Sciences, Bengaluru, India-560089 \\
	${}^{2}$ Department of Physics, Technical University of Munich,
Boltzmannstraße 3, 85748 Garching, Germany\\
${}^{3}$ Department of Mathematics, Technical University of Munich,
Boltzmannstraße 3, 85748 Garching, Germany
}

\date{Received: date / Accepted: date}

\maketitle

\begin{abstract}
We consider the relaxation of an initial non-equilibrium state in a one-dimensional fluid of hard rods. Since it is an interacting integrable system, we expect it to reach the  Generalized Gibbs Ensemble (GGE) at long times for generic initial conditions. Here we show that there exist initial conditions for which the  system does not reach GGE even at very long times and in the thermodynamic limit. In particular, we consider an initial condition of uniformly distributed hard-rods in a box with the left half having particles with a singular velocity distribution (all moving with unit velocity) and the right half particles in thermal  equilibrium. We find that the density profile for the singular component does not spread to the full extent of the box and keeps  moving with a fixed effective speed at long times. We show that such density profiles can be well described by the solution of the Euler equations almost everywhere except at the location of the shocks, where we observe slight discrepancies due to dissipation arising from the initial fluctuations of the thermal background. To demonstrate this effect of dissipation analytically, we consider a second initial condition with a single particle at the origin with unit velocity in a thermal  background. We find that the probability distribution of the position of the unit velocity quasi-particle  has diffusive spreading which can be understood from the solution of the Navier-Stokes equation of the hard rods.
Finally, we consider an initial condition with a spread in velocity distribution for which we show convergence to GGE. Our conclusions are based on molecular dynamics simulations supported by analytical arguments. 

\end{abstract}
\tableofcontents

\section{Introduction}
A classical Hamiltonian many-body system will generally thermalize at long times in the sense that macroscopic obeservables can be described by the Gibbs Ensemble (GE). However, there may exist systems that do not thermalize to GE, because of the existence of macroscopic number of extra conservation laws which restrict their motion in the phase space. 
Such systems are often known as integrable many-body systems, which are believed to thermalize to the Generalized Gibbs Ensemble (GGE)\cite{PhysRevLett.98.050405,Pozsgay_2013,GGE,Vidmar_2016}. They have been realized experimentally in one-dimensional trapped atoms \cite{cradle,doi:10.1126/science.abf0147}. Their non-equilibrium dynamics close to local GGE is described by generalised hydrodynamics (GHD) \cite{doyonyoshimura,PhysRevLett.117.207201,Alba_2021,doyon1,doyon2}.
Integrable systems are very fine-tuned systems in the sense that the smallest of perturbations (which are always present in any experimental setup) can break integrability. However, in the presence of integrability breaking perturbations, it is expected that the system will still remain integrable for short times \cite{cradle,doi:10.1126/science.abf0147,Bastianello_2021,Bastianello_2022}, and so integrable dynamics may still play an improtant role.
Integrable systems are also important from the point of view of studying exact dynamics of systems far from equilibrium. Since they rekindle the hope of obtaining exact solutions to many-body systems out of equilibrium, it is possible to use them to study far from equilibrium states, which cannot be treated using hydrodynamics (for non-integrable case) or generalised hydrodynamics (for integrable case) because hydrodynamics (HD) can only handle states near local equilibrium (or local GGE). 
\par
It is useful to make a distinction between interacting  and non-interacting integrable systems \cite{10.1063/1.5018624}. In non-interacting integrable systems, the quasiparticles move in straight lines at constant velocity.  For example, in one dimensional hard point particle gas, the collisions happen at a point and two particles simply exchange velocity after colliding, and thus the system can be mapped to non-interacting one by interchanging the labels of the two colliding particles after the collision. In the mapped non-interacting problem, the new particles moving with a fixed velocity are called quasi-particles, and they move in straight lines at constant velocities, like in a non-interacting gas. This is not the case for interacting integrable systems. In the hard rod case, a quasiparticle  will have a straight line motion interrupted by sudden jumps (of the size of rod length) owing to the collisions. This effectively leads to dissipation in the hydrodynamics of the hard rod gas, which can get manifested by the spreading of a tagged quasiparticle. On the other hand, for the point particle case, the system has no dissipation term in its hydrodynamics and consequently no spreading in the position of a tagged quasiparticle. Such spreading was studied, from microscopic calculations, by Lebowitz, Percus and Sykes~\cite{LPS} and demonstrated the effect of dissipation. Such dissipation terms appearing as  Navier-Stokes (NS) corrections to the HD equation of hard rods was later established by Spohn~\cite{spohnbook}, Boldrighini and Suhov~\cite{boldighrini2} and recently discussed by Doyon and Spohn~\cite{doyonspohn} and Ferrari and Olla~\cite{ferrari2023}.  Due to the presence of the dissipation term, one generally expects that the hard rod gas would approach to a GGE state starting from a non-equilibrium initial condition.

The question of approach to the GGE state in integrable systems has been widely discussed in the quantum context~\cite{PhysRevLett.98.050405,PhysRevLett.106.140405,PhysRevLett.106.227203,PhysRevLett.127.020501} and the effect of dissipation was demonstrated in the context of evolution of a domain wall in the quantum Heisenberg spin chain \cite{De2018}. However, to the best of our knowledge, this has not been observed for classically integrable systems. Neither has  the effect of the Navier-Stokes correction to the Euler GHD solutions been demonstrated in any study. In the context of the classical system of hard rods, the  questions on evolution and effect of dissipation were addressed in~\cite{doyonspohn} for the specific case of domain wall initial condition. This study demonstrated that the evolution from such initial condition can be very well accurately described by the solution of the Euler GHD equations. Although the corrections  from the Navier-Stokes terms were discussed in ~\cite{doyonspohn}, this  could not be unambiguously established from the numerics. The aims of the present paper are: (i) to study the evolution of non-equilibrium initial states  and see if they approach GGE  at large times; (ii) to demonstrate the effect of dissipation in such an evolution. 

\par
This paper is organized as follows. In Sec.~\ref{sec:2} we define the model and  the different initial conditions used in the study, and summarize the main results.  We investigate the equilibration and the effect of dissipation in Sec.~\ref{sec:3} by comparing the predictions of hydrodynamics with those of MD simulations for different initial conditions. In Sec.~\ref{sec:dis} we provide discussions of our results and conclude. Some details of the calculations are provided in the appendix. 

\section{Model, observables and initial conditions}
\label{sec:2}

We consider a system of $N$ hard rods each of length $a$ and unit mass, moving inside a one dimensional box of size $L$. The rods move with constant velocity in between collisions. Two rods exchange their velocities at collisions with each other whereas at collisions with the walls at $x=0$ and $x=L$ the rods flip their velocities. This implies reflecting boundary conditions at $x=0$ and $x=L$. 
This model with $a\neq 0$ is an example of an interacting integrable system, while for $a=0$, it becomes non-interacting integrable system. 

The microscopic dynamics of hard rods can be mapped to that of hard point particles as 
 follows \cite{Percus,BernsteinPercus,LPS}: Let $x_i,~i=1,2,...,N$ denote the ordered positions ($x_{i}<x_{i+1}-a$ with $x_1>a/2$ and $x_N< L-a/2$) and $v_i,~~i=1,2,...,N$ denote velocities of the rods. For each microscopic configuration $\{x_i,v_i\}$ of hard rods, one can construct a configuration $\{x'_i,v'_i\}$ of hard point particles by removing the inaccessible spaces between rods and, between rods and the walls. More precisely the mapping can be written as 
\begin{align}
x'_i=x_i - (i-1/2)a,~~v'_i=v_i~~i=1,2,...,N, \label{hr-pp-map}
\end{align}
and consequently one has a set of hard point particles moving inside a box of length $L'=L-Na$. The dynamics of hard point particles can be further mapped to non-interacting point particles by the Jepsen mapping~\cite{jepsen1965dynamics}. This mapping has earlier been used to find several analytical results, such as quasiparticle distribution~\cite{LPS}, free expansion problem~\cite{BernsteinPercus} and sound and shock propagation~\cite{Percus}.~
This mapping also allows one to simulate the hard rod dynamics efficiently and accurately. Throughout this paper, we represent  configurations of the point particles by primed variables ($\{x_i',v_i'\}$) and those of the rods by un-primed variables ($\{x_i,v_i\}$).

\par
In this paper we study the evolution of the single particle phase space distribution, $f(x,v,t)$, of the hard rods defined as 
\begin{align}
f(x,v,t)= \left \langle\sum_{i=1}^N\delta(x-x_i)\delta(v-v_i) \right \rangle,
\end{align} 
where $\langle ... \rangle$ denotes an   average over the ensemble of initial conditions corresponding to fixed forms of the initial density profile and single particle velocity distribution. We investigate the possible approach to GGE and the effect of NS terms, for the following three different initial conditions: 
\begin{enumerate}
\renewcommand{\labelenumi}{\Alph{enumi}}
\item  The particles on each of the two halves, $(0,L/2)$ and $(L/2,L)$, are separately distributed uniformly  at a finite density $\rho_{0}=N/L$.  We assign all particles on the left half [$x \in (0,L/2)$] with velocity $v_0$($=1$), while the velocities of all particles on  the right half [$x \in (L/2,L)$] are chosen from a Maxwell distribution with temperature $T=1$ 
\begin{align}
h(v)=\frac{1}{\sqrt{2 \pi}}\exp \left(-\frac{v^2}{2}\right),~~\text{for}~~-\infty \le v \le \infty. \label{p(v)_mx}
\end{align}
In this initial condition, one has two components of the gas -- for the first component,  each particle has velocity $v_0=1$, while in the second (which we call the background particles),  the particles have velocity distributed according to $h(v)$.

\item  Next we consider an initial condition where we first place a particle with a given velocity $v_0$(=1) at the origin. The rest of the box is then filled with particles distributed uniformly in space with a density $\rho_0=N/L$. These background particles  have velocities chosen from the Maxwell velocity distribution $h(v)$.  This case is more analytically tractable than the previous case in (A) and was first studied by Lebowitz, Percus, Sykes (LPS) in \cite{LPS}.  In this case, the initial single particle phase space distribution has the form $f(x,v,t=0)=\delta(x)\delta(v-v_0)+\rho_0h(v)$. \\  

\item  Finally we consider the set-up of free expansion from a half filled box. In this case the rods are uniformly distributed in the left half of the box at a constant density, $2 \rho_0$,  and the velocities are again chosen from the Maxwell distribution, $h(v)$.  The right half of the box is empty. This problem of free expansion was previously investigated in \cite{BernsteinPercus}, where the evolution of various hydrodynamic variables was computed using a microscopic approach and with certain approximations that effectively amount to solving the Euler equations. For both the classical and quantum cases, the free expansion problem for point particles has recently been  studied in the context of entropy growth~\cite{chakraborti,chakrabortiinteracting,pandey2023boltzmann}.
\end{enumerate}

For the three initial conditions mentioned above, we study the evolution of the  density profile, $\rho(x,t)$, and the velocity profile, $u(x,t)$  (or equivalently, the momentum density profile $p(x,t)=\rho(x,t)u(x,t)$), defined as 
 \begin{subequations}
 \begin{align}
     \rho(x,t)&=\int f(x,v,t)dv, \label{rho(x,t)} \\
     u(x,t)&=\frac{1}{\rho}\int vf(x,v,t)dv. \label{u(x,t)} 
 \end{align}
 \end{subequations}
We investigate the effect of dissipation by comparing the profiles obtained from simulation  with those predicted from the solution of Euler GHD. We also check if the system reaches GGE at long times. A simple test for GGE would be to check if the density  and  velocity profiles become time stationary, {\it i.e.}, independent of time and uniform in space.  Note that the velocity distribution is invariant under the integrable dynamics and specifies the GGE. We summarize here our main findings: 
\begin{itemize}
    \item For initial condition (A), we find that the predictions for the evolution of  the densities, from the solution of the Euler GHD, describe the profiles obtained from numerical simulations quite well almost everywhere except at the locations of the shocks where we observe clear discrepancies. These discrepancies appear due to the effect of dissipation described by the NS term. This leads to the width $w(t)$ of the shock growing as $w(t) \sim \sqrt{t}$ at early times, and saturating to a value $w(t \to \infty) \sim \sqrt{N}$ at large times.  Thus the initial density of either of the components ($v_0=1$ and the thermal ones) never become homogeneous over the full system and each of the two components move inside the box with a constant effective speed $v_{\text{eff}}$ (see Eq.~\eqref{veff}) at large times. Hence the system never reaches  GGE.

    \item This spreading is more prominent in the case of initial condition (B). For this case we provide an analytical understanding of the spreading based on the solution of the NS equation. In this case also the system never reaches GGE.

    \item For the case (C) of free expansion we find that the evolution of the density and momentum density profiles is completely described by  Euler GHD and, since the discontinuity in the initial density profile disappears already at very early times,  any effects of the NS corrections are too small to be observed. In this case, the system  at long times evolves to a state consistent with GGE.    
\end{itemize}

\section{Hydrodynamic equations for hard rods  and solution of the Euler equation}
\label{sec:hd-eq}
The hydrodynamic equation for the single particle phase space distribution $f(x,v,t)$ for the hard rod gas is given by~\cite{doyonspohn}: 
 \begin{subequations}
\label{hdeq}
\begin{align}
\partial_tf(x,v,t)&+\partial_x\left(v_{\rm eff}(x,v,t)f(x,v,t)\right)=\partial_x\mathcal{N}(x,v,t), ~~ 
\text{where}~~\\v_{\rm eff}(x,v,t)&=\frac{v-a\rho(x,t) u(x,t)}{1-a\rho(x,t)}, \label{NS-equation}\\
\mathcal{N}(x,v,t)&=\frac{a^2}{2(1-a\rho(x,t))}\int dw~|v-w|~\left(f(x,w,t)\partial_xf(x,v,t)-f(x,v,t)\partial_xf(x,w,t)\right) 
\end{align}
\end{subequations}
and $\rho(x,t)$,  $u(x,t)$ are given in 
Eq.~\eqref{rho(x,t)} and \eqref{u(x,t)}. The term $\partial_x \mathcal{N}$ represents the NS correction to the Euler equations~\cite{doyonspohn,boldighrini2}.

We now discuss the solution of the Euler equation 
\begin{align}
\partial_tf+\partial_x(v_{\rm eff}f)&=0, \label{EGHD}
\end{align}
for general initial condition. As shown previously~\cite{BernsteinPercus,Percus}, the Euler equation for hard rods can be solved exactly for general initial conditions by mapping it to a non-interacting point particle problem. 
For completeness, we show below how the mapping to the non-interacting Euler equation can be obtained using the GHD approach. 
 For this one defines a new function,
 \begin{equation}
     \label{f-to-f^0}
     f^0(x',v,t)=\frac{f(x,v,t)}{1-a\rho(x,t)},
 \end{equation}
 where 
 \begin{align}
    x'&=x-aF(x,t),~~
F(x,t)=\int_{B}^x\rho(y,t)dy,
 \end{align} 
 and $B$ is the position of the left end of the container in which the hard rod fluid is contained.  Note that $F(x,t)$ is the cumulative density corresponding to $\rho(x,t)$. We observe that 
 \begin{equation}
   f^0(x',v,t)~dx'dv = f(x,v,t)~dxdv,  \label{jacob}
 \end{equation} 
 which implies that $f^0(x',v,t)$ is also a phase space distribution function.
 We now show that $f^0(x',v,t)$  satisfies  the  Liouville equation of free ballistic particles and hence describes the single particle phase space distribution of the point particles. The first step towards this demonstration~\cite{doyon1} is to define the function 
 \begin{equation}   f_n(x,v,t)=\frac{f(x,v,t)}{1-a\rho(x,t)}. 
 \end{equation}
 Using Eq.~\eqref{EGHD} and the relation $\partial_t \rho=-\partial_x (\rho u)$ [for the fields defined in Eqs.~(\ref{rho(x,t)},\ref{u(x,t)})], it readily follows that  
 \begin{equation}
     \partial_tf_n+v_{\rm eff}\partial_xf_n=0.  \label{f_n}
 \end{equation}
 Now, from Eq.~\ref{f-to-f^0}, we have
 \begin{equation}
     f_n(x,v,t)=f^0(x-aF(x,t),v,t).  \label{eq:f_n-1}
 \end{equation}
Taking the time derivative with respect to $t$ on both sides of  Eq.~\eqref{eq:f_n-1} and using $\partial_tF(x,t)+\rho u=0$ one finds
 \begin{equation}
 \label{partial_tf_n}
\partial_tf_n(x,v,t)=\partial_tf^0(x',v,t)+a\rho(x,t) u(x,t)\partial_{x'}f^0(x',v,t).
 \end{equation}
On the other hand taking derivative with respect to $x$ on both sides of  Eq.~\eqref{eq:f_n-1} one has 
 \begin{equation}
 \label{partial_xf_n}
     \partial_xf_n(x,v,t)=(1-a\rho)\partial_{x'}f^0(x',v,t).
 \end{equation}
 Inserting the forms  from Eq.~
\ref{partial_tf_n} and Eq.~\ref{partial_xf_n} in  Eq.~\ref{f_n}, one finds that the  phase space distribution function $f^0(x',v,t)$ satisfies the Liouville equation for the non-interacting  particles 
 \begin{equation}
     \partial_tf^0+v\partial_{x'}f^0=0.
     \label{eghd-pp}
 \end{equation}
This equation can be easily solved for arbitrary time and any initial condition $f^0(x',v,0)$. For example on the infinite line one has $f^0(x',v,t)=f^0(x'-vt,v,0)$ while in the box one has to solve the single particle problem with repeated collisions with the walls \cite{BernsteinPercus,chakraborti}. From  $f^0(x',v,t)$ one can find the solution for the phase space distribution $f(x,v,t)$ of  hard rods. To get the solution explicitly, we first note from Eq.~\eqref{f-to-f^0} that
 \begin{equation}
 \label{rho^0-to-rho}
     \rho(x,t)=\frac{\rho^0(x',t)}{1+a\rho^0(x',t)},~~~
     \text{where}~~\rho^0(x',t)=\int dv~f^0(x',v,t).
 \end{equation}
Hence inverting Eq.~\eqref{f-to-f^0} and using  Eq.~\eqref{rho^0-to-rho}, one finds
 \begin{equation}
     \label{eq:48} f(x,v,t)=\frac{f^0(x',v,t)}{1+a\rho^0(x',t)}.
 \end{equation}
The variable transformation $x \rightarrow x'$ can be inverted  as 
\begin{align}
 x=x'+aF^0(x',t),   \label{map-x'-to-x}
\end{align} 
using $F(x,t)=F^0(x',t)$ which can be shown easily from Eq.~\eqref{jacob}.

While, as demonstrated above, the Euler equation can be solved exactly, it is difficult to solve the NS equation~\eqref{NS-equation} for arbitrary initial conditions. We expect that the difference between the solutions of the Euler and the NS equations are large at places where the spatial derivative of the Euler solution is large. 

\section{Results from numerical simulations for the three initial conditions}

\subsection{{Initial condition A}}
\label{sec:3}
In this case the initial condition can be written explicitly as
\begin{align}
f(x,v,0) &= g(x,0)\delta(v-1) + f_b(x,v,0),~~\text{with}~~f_b(x,v,0)=\rho_b(x,0)h(v),~~\text{for}~\frac{a}{2} \le x \le L-\frac{a}{2},
\end{align}
where $h(v)$ is given in Eq.~\eqref{p(v)_mx} and 
\begin{align}
g(x,0) &= \rho_0 ~\Theta(x-a/2)\Theta(L/2-x) \label{g(x,0)}\\
\rho_b(x,0) &= \rho_0 ~\Theta(x-L/2) \Theta(L-a/2 -x), \label{rho_b(x,0)}
\end{align}
with $\rho_0=\frac{N}{L-a}$ and $\Theta(x)$ being Heaviside theta function. Note that we will be working in the thermodynamic limit $N \to \infty$ $L \to \infty$ such that $N/L \to \rho_0$.

\begin{figure}[hbt!]
\centering
\begin{subfigure}{.45\textwidth}
\centering
\includegraphics[width=1.0\linewidth]{{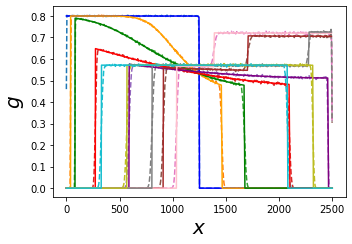}}
\caption{}
\end{subfigure}%
\begin{subfigure}{.45\textwidth}
\centering
\includegraphics[width=1.0\linewidth]{{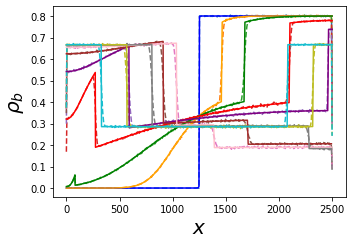}}
\caption{}
\end{subfigure}
\caption{{\bf Comparing solution of Euler equation with MD simulation for initial condition A}: Plot comparing the solution of the Euler equation with those of molecular dynamics for (a) the density of $v=1$ particles, denoted by $g(x)$ and (b) the density of background particles, denoted by $\rho_b(x)$. Dashed lines are MD simulations and solid lines are solutions of Euler equation.   We have taken times $t=0$ (dark blue), $t=40$ (orange), $t=80$ (green), $t=160$ (red), $t=240$ (violet), $t=320$ (brown), $t=400$ (pink), $t=480$ (grey), $t=560$ (mud green) and $t=640$ (cyan). We see that even for very long time like $t=640$, the profile obtained from MD does not relax to GGE, i.e., it does not become uniform. We also see that there is a discrepancy between MD and Euler solutions at the shock front due to dissipative effects. We have taken length of the box $L=2500$, total number of particles $N=2000$ and length of rod $a=1.0$. We have performed ensemble averaging over $5000$ realizations while doing MD.}
\label{fig:1}
\end{figure}

As discussed in the previous section, solution to the Euler equation can be obtained by mapping to point particles. It is easy to show that initial phase space density $f^0(x',v,0)$ also has two components, the special component with velocity $v=1$ and the background particles having velocity distributed according to Maxwell distribution. It is given explicitly as 
\begin{align}
f^0(x',v,0) &= g^0(x',0)\delta(v-1) + f_b^0(x,v,0),~~\text{with}~~f_b^0(x,v,0)=\rho_b^0(x,0)h(v),~~\text{for}~0 \le x' \le L',
\end{align}
where $L'=L-Na$ and 
\begin{align}
g^0(x',0) &= \frac{\rho_0}{1-a\rho_0} ~\Theta(x')\Theta(L'/2-x') \label{g^0(x,0)}\\
\rho_b^0(x',0) &= \frac{\rho_0}{1-a\rho_0}  ~\Theta(x'-L'/2) \Theta(L' -x'). \label{rho_b^0(x,0)}
\end{align}
The evolution of $f^0$ for any  arbitrary initial distribution $f^0(x',v,0)=\varrho(x') p(v)$ is given by [see Appendix~A of \cite{chakraborti}]
\begin{align} 
\begin{split}
f^0(x',v,t) =  \int_0^{L'} dy ~\varrho(y) \int_{-\infty}^{\infty}du ~p(u) \sum_{n=-\infty}^{\infty} [ & \delta(x'-y-ut+2nL') 
 \delta(v-u)  \\
 &+ \delta(x'+y+ut-2nL') \delta(v+u) ], 
 \end{split}\label{f_eq1} 
\end{align}
Since the Euler Equation \eqref{eghd-pp} for point particles is linear, the distribution $f^0(x',v,t)$ at time $t$ still can be written as a sum of two components as 
\begin{align}
    f^0(x',v,t)= g^0(x',t)\delta(v-1) + f_b^0(x',v,t). \label{sol:f^0(t)}
\end{align}
The first term in this expression can be obtained by putting $\varrho(y)=g^0(y,0)$ and $p(u)=\delta(u-1)$ and performing  the integration over $v$, yielding
\begin{align}
    \begin{split}
    g^0(x',t)=\frac{\rho_0}{1-a\rho_0}\sum_{n=-\infty}^{\infty}(\Theta(x'-t+2nL')-\Theta(x'-t+2nL'-L'/2)\\+\Theta(-x'-t+2nL')-\Theta(-x'-t+2nL'-L'/2)),
    \end{split}
    \label{f_eq2}
\end{align}
where, recall $L'=L-Na$.
For the background component we set $\varrho(y)=\rho_b^0(y,0)$ and $p(u)=h(u)$ to get:
\begin{align} \nonumber
\rho_b^0(x',t) &  =  \int_{-\infty}^{\infty}dv f_b^0(x',v,t), \\ \nonumber
&= \frac{\rho_0}{1-a\rho_0} \frac{1}{\sqrt{2\pi T}} \sum_{n=-\infty}^{\infty} \int_{L'/2}^{L'} dy \int_{-\infty}^{\infty}dv~ e^{-v^2/2T} \left[ \delta(x-y-vt+2nL) 
+ \delta(x+y-vt-2nL) \right], \\ \nonumber
&= \frac{\rho_0}{1-a\rho_0} \frac{1}{\sqrt{2\pi T}} \sum_{n=-\infty}^{\infty} \int_{L'/2}^{L'} dy \frac{1}{t} \left[ \exp \left\{ \frac{-(2nL'+x-y)^2}{2Tt^2} \right\} 
+ \exp \left\{ \frac{-(2nL'-x-y)^2}{2Tt^2} \right\} \right], \\
&= \frac{\rho_0}{2(1-a\rho_0)}  \sum_{n=-\infty}^{\infty} \left[\text{erf}\left( \frac{x+L'/2 +(2n-1)L'}{\sqrt{2T}t} \right) - \text{erf}\left( \frac{x-L'/2 +(2n-1)L'}{\sqrt{2T}t} \right) \right]. \\
\label{eq_density_field}
&= \frac{\rho_0}{2(1-a\rho_0)}  \sum_{n=-\infty}^{\infty} \left[\text{erf}\left( \frac{x-L'/2 +2nL'}{\sqrt{2T}t} \right) - \text{erf}\left( \frac{x+L'/2 +2nL'}{\sqrt{2T}t} \right) \right].
\end{align}
Using the Poisson resummation formula, this can  be rewritten in the alternative series form: 
\begin{align}
\label{eq_density_field2}
\rho_b^0(x',t)=\frac{\rho_0}{2(1-a\rho_0)}+\frac{2\rho_0}{(1-a\rho_0)}\sum_{k=1}^{\infty}\frac{1}{\pi k}\cos{\left(\frac{k\pi(L'-x)}{L'}\right)}\sin{\frac{k\pi}{2}}e^{-\frac{\pi^2k^2t^2}{2{L'}^2}}.
\end{align}
Note that shifting the origin to $L'/2$ ({\it i.e.}, $x' \to z'=x'-L'/2$) and taking $L' \to \infty$, one obtains the solution of Euler GHD on the infinite line as 
\begin{align}
\begin{split}
g^0(z',t)&=\frac{\rho_0}{1-a\rho_0}\Theta(z'-t), \\
\rho_b^0(z',t) &= \frac{\rho_0}{2(1-a\rho_0)} \left [1+\text{erf} \left( \frac{z'}{\sqrt{2T} t} \right) \right],
\end{split}
\label{sol-EGHD-inf-line}
\end{align}
where $T=1$. The corresponding densities of the hard rods for the two components, respectively $g(x,t)$ and $\rho_b(x,t)$, can be obtained using the inverse mapping Eq.~\ref{rho^0-to-rho} and Eq.~\eqref{map-x'-to-x} along with $\rho^0(x',t)=g^0(x',t)+\rho_b^0(x',t)$.  
 We  show in Fig.~\ref{fig:1}  the evolution of $g(x,t)$ and $\rho_b(x,t)$ obtained from the solutions of the Euler GHD equation as well as  the results from direct MD simulations.  

\begin{figure}[hbt!]
\centering
\includegraphics[width=1.0\linewidth]{{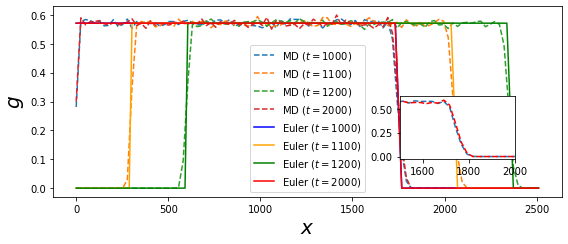}}
\caption{{\bf Lack of thermalization to GGE}: Plot of $g(x,t)$ {\it vs.} $x$ at different (late) times.  The dashed lines are obtained from molecular dynamics and the solid lines represents the solutions of the Euler equations. We observe that the profiles at $t=1000,~1100$ and $t=1200$ have moved by a displacement $\Delta x \approx 300$, implying $v_{\rm eff} \approx 3$, in agreement with Eq.~\eqref{veff}.   The width of the pulse at the times $t=1000,2000$ are the same, thus indicating that it saturates and the whole profile does not become uniform, i.e., it does not relax to a GGE form. The inset shows a zoom of the  shock at the two times $t=1000$ and $t=2000$, where we see that its width has saturated. We chose $t=1000$ and $t=2000$ as  times for which the  profiles coincided for the particular parameter values (i.e., $a,N,L$). Here $N=2000$, $L=2500$, $a=1$ and ensemble averaging over $100$ realizations were performed while doing MD. The values of $g_1,\rho_b,\bar{\rho}_b$ and $v_{\rm eff}$ agree with the  predictions in Sec.~\ref{sec:3}).  \label{fig:3}}
\label{large-time-sol}
\end{figure}

For the solutions of the Euler GHD, we make the following observations:  
\begin{itemize}
    \item[a.]  There is always a shock at the front of the density profiles for both the components. On the infinite line, the shocks for the two components move in opposite directions. Note that the density profiles $g^0(x',t)$ and $\rho^0(x',t)$ in the point particle gas evolve independently of each other.  Consequently, $g^0(x',t)$ will move with constant speed $v_0=1$ keeping the initial shape unchanged, {\it i.e.}, with two discontinuities at $L/2$ separation. Hence the total density, $\rho^0(x',t)=g^0(x',t)+\rho^0_b(x',t)$ will also have discontinuities. Consequently, the density profiles $g(x,t)$ and $\rho(x,t)$ of the hard rods, obtained through the transformation in Eq.~\eqref{rho^0-to-rho} also exhibits discontinuities, {\it i.e.,} shocks.

    \item[b.] At early times the evolution of these density profiles correspond to that on an infinite line and can be described by $g(x,t)$ and $\rho_b(x,t)$ obtained after transforming the solutions given in Eqs.~\eqref{sol-EGHD-inf-line} for the Euler equation of the point particles.

    \item[c.] At later times, each component of the gas gets reflected from the walls of the box which  are described, in the point particle picture, by various terms in the series in Eq.~\eqref{f_eq2} and Eq.~\eqref{eq_density_field}. 

    \item[d.] At the longest times both density profiles $g(x,t)$ and $\rho_b(x,t)$ stop broadening further and settle to piece-wise flat profiles which move between the walls with some constant effective velocity $v_{\rm eff}$ (see Fig.~\ref{large-time-sol}). The details of this solution will be discussed below. Since the density profile does not become time stationary  even at the largest times, this indicates that  for initial condition A, the hard rod system will never reach a GGE state ( which should be time stationary). 
\end{itemize} 

\begin{figure}[hbt!]
\centering
\begin{subfigure}{.45\textwidth}
\centering
\includegraphics[width=1.0\linewidth]{{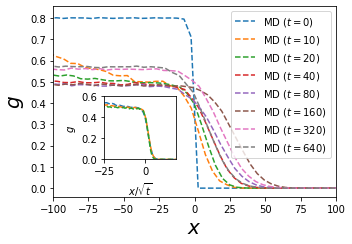}}
\caption{}
\end{subfigure}%
\begin{subfigure}{0.45\textwidth}
\centering
\includegraphics[width=1.0\linewidth]{{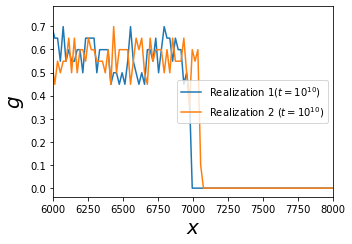}}
\caption{}
\end{subfigure}%
\caption{{\bf (a) Time dependence of the width and (b) the fluctuation in the location of the shock}:  (a) This shows the structure of the shock for the initial condition A at different times for a given system size ($N=2000$, $L=2500$). The curves have been shifted so that the shock fronts for all the curves coincide. While doing MD, ensemble averaging over $5000$ initial microstates were performed. We see that there is trend of increasing width with time while in the inset (which shows curves for $t=20,40,80$), we see that there is a scaling collapse in the variable $x/\sqrt{t}$ for short times when the $v=1$ pulse does not know about the boundaries of the system and hence behaves like it is in an infinite system. We explain this $\sqrt{t}$ dependence in Appendix \ref{app:B}.   (b) This shows plot of $g(x,t)$ for two different realizations (microstates) for initial condition A. We see that in a single realization  the shock remains sharp, while the positions of the shock front in two realizations are different. Consequently, ensemble averaging will lead to smearing of the shock and thus is necessary to observe dissipation.  Here we have chosen $N=8000$, $L=10000$.  
}
\label{spreading-in-ICA}
\end{figure}

\begin{figure}[hbt!]
\centering
\includegraphics[width=1.0\linewidth]{{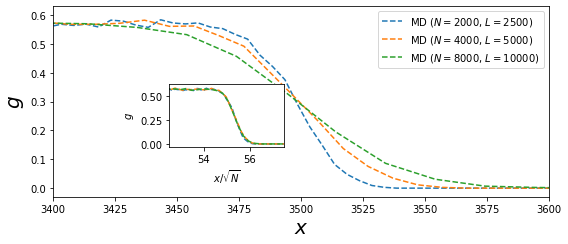}}
\caption{{\bf System size dependence of the shock width for initial condition A}: This  shows the structure of the shock at late times (when the width saturates) for the initial condition A for different system sizes. For all the curves we have chosen $t=10000$ which is much longer than the time at which the width of the pulse $g(x,t)$ and that of the shock saturates. Even after this long time, the curve $g(x,t)$ does not become uniform, i.e., it does not relaxes to GGE.  The curve has been shifted so that the shock fronts for all the three curves coincide. We see that the shock broadens with system size, while in the inset, we see that there is a scaling collapse in the variable $x/\sqrt{N}$, thus showing that the shock broadens with the system size as $\sqrt{N}$. 
In this case ensemble averaging over $500$ realizations was  performed.}
\label{sharp}
\end{figure}

While we see a very good overall agreement between Euler solution and MD simulations, there are clear differences. If we zoom near the shocks  in Figs.~\ref{fig:1}a and \ref{fig:1}b, we notice that the simulation data for the hard rod density profile $g(x,t)$ (dashed lines) shows a slight discrepancy with the Euler prediction. 
One observes similar discrepancy for $\rho_b(x,t)$ also.  The simulated profiles display spreading at the locations of the shock in the Euler solutions. This is demonstrated in Figs.~(\ref{spreading-in-ICA}(a),\ref{sharp}), where the density profiles are zoomed near the shock location after shifting appropriately so that the shock positions coincide. This spreading is a signature  of the dissipation characterised by the Navier-Stokes term in Eq.~\eqref{hdeq}.
We observe that the width of the shock increases with time 
and scales as $\sqrt{t}$ as can be seen from Fig.~\ref{spreading-in-ICA}(a) where profiles for different times collapse under the scaling of $x$  by $\sqrt{t}$.  Microscopically the spreading originates from the fluctuations in the number of the background rods (having Maxwell velocity distribution) that a shock, of the Euler solution for $g(x,t)$, encounters till time $t$. This fluctuations arise from the fluctuations in the initial conditions. For a given initial configuration of the positions and velocities of the rods, the shock remains sharp and does not widen. However, the place at which the shock appears at a given time fluctuates from one initial microstate  to another (see Fig.~\ref{spreading-in-ICA}(b)). This happens because the  number of background rods that the special rods encounters is different for different initial microscopic configurations. Hence, on an average the shock widens. At small times, these fluctuations are independent as the rods have not realised the presence of   boundaries of the box. The  $\sqrt{t}$ growth at small times can be explained by considering the evolution of the density profile starting from initial condition A on an infinite line which is done in Appendix \ref{app:B}. 
 
 The early time growth of the width of the shock  stops after some time and saturates to a $O(\sqrt{N})$ value as demonstrated in Fig.~\ref{sharp}. As time progresses the rods move back and forth inside the box and consequently, the fluctuations in the number of background rods inside the region of the special rods (having velocity $v_0$) do not remain independent and get correlated. Consequently, the spreading of the shock cannot continue to grow as $\sqrt{t}$ and saturates to the observed $O(\sqrt{N})$ value. Thus even in the thermodynamic limit the pulse $g(x,t)$ does not spread to the full extent of the system and remains in the shape of a rectangular pulse that keeps on moving back and forth inside the box. Consequently, the total density profile of the rods does not become homogeneous and stationary as one would expect in a GGE state. This implies that the a hard rod system inside a box, starting from initial condition A, does not reach the GGE state even in the thermodynamic limit.

\paragraph{Euler solution in $t\rightarrow\infty$ limit:}
{We now find the solution of the Euler equation in the $t\rightarrow\infty$ limit. Recall that in the initial condition A, the rods are uniformly distributed in each half with density $\rho_{0}$. The rods on the left half have velocity $v=1$ and those on the right half have velocities distributed according to Eq.~\eqref{p(v)_mx}. Using Eq.~\ref{rho^0-to-rho}, one maps this hard rod system to a point particle gas with uniform density $\rho^0(x,0)=\frac{\rho_{0}}{1-a\rho_{0}}$ inside a smaller box of size $L'=L-Na$. The velocity distribution remains unchanged as in the hard rod gas, {\it i.e.}, $\delta(v-1)$ in the left half and $h(v)$ in the right half. In the point particle gas the component with velocity $v=1$ moves without changing its shape  whereas the particles on the right half (called the background particles) perform free expansion, ignorant of the $v=1$ particles since the gas is non-interacting. At long times, the background point particles will expand into the full box of length $L'$ and become uniform with density half of their initial density, {\it i.e.}, $\rho^0(x,t \to \infty) = \frac{\rho^0(x,0)}{2} =\frac{\rho_{0}}{2(1-a\rho_{0})}$. Thus at long times, one would observe the initial density pulse of the special point particles with velocity $v=1$ moving in the uniform background of thermal particles (with Maxwell velocity distribution). Hence, at any instant, the total density profile has two regions: a uniform high density region where the $v=1$ pulse is present (we call it the pulse region) and  a uniform low density region in the remaining part of the box. Thus, the total density profile in the long time limit becomes piece-wise uniform which we now proceed to compute.

Let us denote the value of the density of the $v=1$ particles inside the pulse by $g_1'$, in the point particle picture, and by $g_1$ in the hard rod picture. Similarly, we denote the density of the background particles inside the pulse region by $\rho_b'$  and $\rho_b$, respectively, in the point particle and hard rod pictures. We also denote the density of background particles outside the pulse region  by $\bar{\rho}_b'$ and $\bar{\rho}_b$, once again, in the point particle  and hard rod pictures, respectively.
The total density of point particles in the pulse region is $g_1'+\rho_b'$ where $g_1'=\frac{\rho_{0}}{1-a\rho_{0}}$ and $ \rho_b'= \frac{\rho_{0}}{2(1-a\rho_{0})}$. Hence the total density there is $\frac{3\rho_{0}}{2(1-a\rho_{0})}$. The density outside the pulse region is  given by $\bar{\rho}_b'=\frac{\rho_{0}}{2(1-a\rho_{0})}$. Now using the inverse mapping in Eq.~\eqref{rho^0-to-rho} along with Eq.~\eqref{map-x'-to-x}, one gets the late time densities in the hard rod picture. The density outside the pulse region is given by 
\begin{equation}
    \bar{\rho}_b=\frac{\rho_{0}}{2-a\rho_{0}},
\end{equation}
and the total density inside the pulse region is given by
\begin{equation}
    g_1+\rho_b=\frac{3\rho_{0}}{2+a\rho_{0}}. \label{rho_b0-g_1-1st-rel}
\end{equation}
To find individual values of $g_1$ and $\rho_b$ we use the conservation of the number of background particles 
\begin{align}
\rho_b \times L_1 + \bar{\rho}_b \times (L-L_1)=\frac{N}{2}, \label{back-part-cons}
\end{align}
where $L_1$ is the length of the pulse region at late times and $L-L_1$ is the length of the region outside the pulse. It is easy to see that  $L_1=\frac{N}{2g_1}$. Dividing both sides of Eq.~\eqref{back-part-cons} by $N$, we get 
\begin{equation}
    \frac{\rho_b}{2g_1}+\bar{\rho}_b\Big(\frac{1}{\rho_{0}}-\frac{1}{2g_1}\Big)=\frac{1}{2}. \label{rho_b0-g_1-2nd-rel}
\end{equation}
Solving Eq.~\eqref{rho_b0-g_1-1st-rel} and Eq.~\eqref{rho_b0-g_1-2nd-rel}, we finally get 
\begin{equation}
    g_1=\frac{2\rho_0}{2+a\rho_0},
\end{equation}
\begin{equation}
    \rho_b=\frac{\rho_{0}}{2+a\rho_{0}}.
\end{equation}
} 
The effective velocity with which the quasiparticles with $v=1$ move at late times can be computed easily. The total density at late times in the pulse region is $\rho=g_1+\rho_b=\frac{3\rho_0}{2+a\rho_0}$. The velocity field $u$ in the pulse region at late times is given by $u=\frac{g_1}{g_1+\rho_b}=\frac{2}{3}$. The effective velocity is thus:
\begin{equation}
\label{veff}
    v_{\text{eff}}=\frac{v_0-a\rho u}{1-a\rho}=\frac{2-a\rho_0}{2-2a\rho_0}.
\end{equation}
 In our MD simulations in Figs.~(\ref{fig:1},\ref{fig:3}), we have taken $\rho_{0}=4/5$,  $a=1$. Plugging these values into the expressions above, we get $\bar{\rho}_b=\frac{2}{3}$, $\rho_b=\frac{2}{7}$, $g_1=\frac{4}{7}$ and $v_{\rm eff}=3$. We have verified that these values match with our MD results at long times in Fig.~\ref{fig:3}. Note that $v_{\rm eff}$ is the late-time speed of quasiparticles with bare velocity $v=1$.

\subsection{Initial condition B}
\label{initialconditionB}
\begin{figure}
\centering
\begin{subfigure}{.45\textwidth}
\centering
\includegraphics[width=1.0\linewidth]{{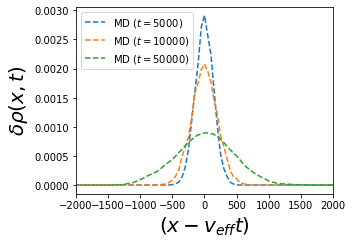}}
\caption{}
\end{subfigure}%
\begin{subfigure}{.45\textwidth}
\centering
\includegraphics[width=1.0\linewidth]{{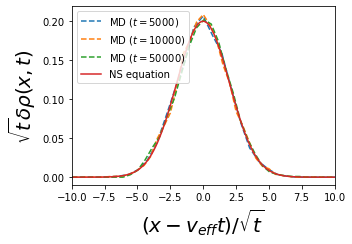}}
\caption{}
\end{subfigure}%
\caption{{\bf Verifying NS equation for LPS-like initial condition}: (a) This figure compares the results from MD simulation for the evolution of the  density profile with those obtained from the solution of the  NS equation, for the LPS-like initial condition.  (b) We show the plot in terms of the scaling variables. We see that there is a good scaling collapse, and a nice agreement with the solution of NS equation.
We have taken, $N=2\times 10^6$, $L=2.5\times 10^6$, $a=1.0$, $v_0=1.0$ (and so $\mu(v_0)\approx 1.0$). For MD, ensemble averaging has been done over $10000$ realizations. The times considered are much before the pulse hits the boundary of the box, hence the system is effectively infinite. \label{fig:LPS}}
\end{figure}

In this case there is a special rod at the origin (middle of the box) with a fixed velocity $v_0=1$ and the two halves of the box on either side of the special particle are  initially filled uniformly  by hard rods.  The velocities of all rods, except the special one,  are distributed according to the Maxwell distribution $h(v)$ given in Eq.~\eqref{p(v)_mx}. This initial condition was studied  by Lebowitz, Percus and Sykes (LPS) in \cite{LPS}. The initial single particle phase space density is 
$f(x,v,t=0)=\delta(v-v_0)\delta(x)+\rho_0h(v)$. Since the initial distribution of the background rods are already in equilibrium, it does not change with time. However, the phase space distribution of the special rod (of velocity $v_0=1$) will change with time. At the Euler level the special rod moves ballistically with an effective velocity $v_{\rm eff} = \frac{v_0}{1-a\rho_0}$. Hence the Euler solution (for an infinite box) is given by  $f(x,v=v_0,t) = \delta(v-v_0)\delta(x-v_{\rm eff}t)$. However, by obtaining the exact microscopic solution of the problem in the thermodynamic limit and by performing an ensemble average, LPS showed that $f(x,v_0,t)$ spreads diffusively, along with a drift with velocity $v_{\rm eff}$ \cite{LPS}, i.e, at long times one has the form $f(x,v_0,t)=\delta(v-v_0) \delta \rho(x,t)$. They also obtained an explicit expression of the diffusion constant.  The  results of LPS were used in \cite{doyonspohn} to compute the current-current correlation and thus the Navier-Stokes (NS) term using the Green-Kubo formula.  In Fig.~\ref{fig:LPS}a we present simulation results for  $\delta \rho(x,t)$ which displays the   spreading predicted by LPS.    We observe that the spreading of the distribution increases with $t$ and the data for different time collapse into a single function under scaling of  space by $\sqrt{t}$, as shown in Fig.~\ref{fig:LPS}b. This implies that the spreading grows with time as $\sqrt{t}$ [at late times it saturates  in a  finite box, due to the same reason for saturation in case (A)]. 

The origin of the growth of the width of the distribution at early times can be understood heuristically from a microscopic computation of the fluctuations of particle number as follows.
Let $\mathcal{N}_t$ be the number of particles in the interval $[0,x_t]$, where $x_t$ is the position of the quasiparticle (special rod) with $v=v_0=1$ at time $t$. In the corresponding point particle picture the special particle, with velocity $v=v_0=1$, would move by a distance $v_0t$ in time $t$. Hence, the position of the rod with velocity $v=v_0=1$ is
\begin{equation}
    x_t=v_0t+a\mathcal{N}_t.
\end{equation}
where $\mathcal{N}_t$, in the point particle picture, is the number of point particles that the special particle has crossed during its evolution, starting from the origin to the position $v_0 t$ at time $t$. The number $\mathcal{N}_t$ fluctuates from one realisation to another in an ensemble of initial conditions, and the fluctuation is proportional to $\sqrt{\langle\mathcal{N}_t\rangle}$. The spread in $f(x,v_0,t)$ will also be proportional to the fluctuations, i.e., to $\sqrt{\langle\mathcal{N}_t\rangle}$. On an infinite line with uniform background of thermal particles, $\langle\mathcal{N}_t\rangle$ grows linearly as $t$ which  thus leads to  the  $\sqrt{t}$ growth of the width in the distribution function. In a finite box, $\langle\mathcal{N}_t\rangle$ cannot grow without bound, because the number of particles in the box is finite.  
 On the hydrodynamic scale,  the $\sqrt{t}$  spreading arises due to the Navier-Stokes terms in Eq.~\eqref{NS-equation} and we  will now demonstrate  this by  obtaining a analytic solution of the Navier-Stokes equation~\eqref{NS-equation} on the infinite line. For this we make the  ansatz: 
\begin{equation}
    f(x,v,t)=\delta(v-v_0)\delta\rho(x,t)+\rho_0h(v),
\end{equation}
where $h(v)$ is given in Eq.~\eqref{p(v)_mx}. 
This ansatz is motivated by the fact that the number of particles with a given velocity is conserved, and that the distribution of the background rods does not change with time. 
Plugging the ansatz into the Navier-Stokes equation, we get the following drift-diffusion equation for $\delta\rho(x,t)$ after ignoring the non linear terms proportional to $(\delta\rho)^2$:
\begin{equation}
    \partial_t(\delta\rho)+v_{\rm eff}\partial_x(\delta\rho)=\frac{n\mu(v_0)a^2}{2}\partial_x^2(\delta\rho),
\end{equation}
where $v_{\rm eff}=\frac{v_0}{1-a\rho_0}$, $n=\frac{\rho_0}{1-a\rho_0}$, $\mu(v_0)=\int dv|v-v_0|h(v)$. 
The solution of this for the LPS-like initial condition is given by:
\begin{equation}
    \delta\rho(x,t)=\frac{1}{\sqrt{2\pi na^2\mu(v_0)t}}e^{-\frac{(x-v_{\rm eff}t)^2}{2a^2n\mu(v_0)t}},
    \label{lps-soln}
\end{equation}
which is exactly the  solution that was obtained by LPS from a completely microscopic analysis~\cite{LPS}.
In Fig.~\ref{fig:LPS}b, we  verify that the expression in Eq.~\eqref{lps-soln} agrees with the MD simulation results.  
Our numerical results thus provide a 
direct demonstration of an observable  effect of the NS terms in the hydrodynamic equations.

 \subsection{Euler vs MD for initial condition C}
\begin{figure}
\centering
\begin{subfigure}{.45\textwidth}
\centering
\includegraphics[width=1.0\linewidth]{{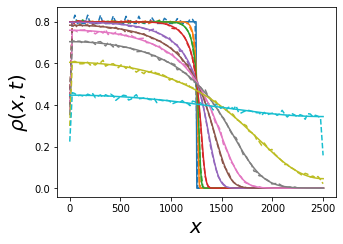}}
\caption{}
\end{subfigure}%
\begin{subfigure}{.45\textwidth}
\centering
\includegraphics[width=1.0\linewidth]{{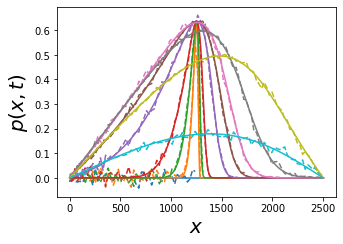}}
\caption{}
\end{subfigure}%
\caption{{\bf Comparing solution of Euler equation with MD simulation for initial condition C}: Plots of the density and momentum profiles and comparison of the exact solution of Euler equation (solid lines) and the profiles obtained from MD (dashed lines) for the free expansion problem. We have shown for times $t=0$ (dark blue), $t=10$ (orange), $t=20$ (green), $t=40$ (red), $t=100$ (violet), $t=150$ (brown), $200$ (pink), $t=300$ (grey), $t=500$ (muddy) and $t=1000$ (light blue). We have taken $N=1000$, $L=2500$ and averaged over $100$ realizations.  \label{fig:6}}
\end{figure}

Finally we consider the free expansion set up in which the  $N$ hard rods are initially confined to the left half of the box of size $L$ and distributed uniformly in space with density $\rho_0=N/L$.  The velocities of the rods are drawn from the Maxwell distribution $h(v)$ in Eq.~\eqref{p(v)_mx}. As in the previous cases, we have hard reflecting walls at  $x=0$ and $x=L$. 
We now follow the same approach  outlined in  Sec.~\eqref{sec:3},  to obtain a solution of the Euler equation for this initial condition, via the mapping to hard point gas.  
The solution in the point particle picture is similar to that obtained in \cite{chakraborti}, with the density given by:
 \begin{equation}
     \rho^0(x',t)=\frac{\rho_0}{1-a\rho_0}+\frac{4\rho_0}{1-2a\rho_0}\sum_{k=1}^{\infty}\frac{1}{\pi k}\cos{\Big(\frac{k\pi x'}{L-Na}\Big)}\sin{\Big(\frac{\pi k(L-2Na)}{2(L-Na)}\Big)}\exp{\Big(\frac{-\pi^2k^2Tt^2}{2(L-Na)^2}\Big)},
 \end{equation}
where $\rho_0=N/L$ and $T=1$. In the $a \to 0$ limit, the above expression of the density profile $\rho^0(x,t)$ matches with those obtained in \cite{chakraborti}. Using the inverse mapping in Eq.~\eqref{rho^0-to-rho}, the density profile $\rho(x,t)$ of the rods can be found, where recall 
 $x=x'+aF^0(x',t)$ and the  cumulative density profile,  $F^0(x',t)$, can be computed from $\rho^0(x',t)$. In Fig.~\ref{fig:6}(a), we  compare the theoretically computed profiles of the rods at different times with the density profiles obtained from MD simulation, and we  observe excellent agreement. From this plot, we observe that with increasing time the density profile of the rods spreads to the right half of the box in a monotonic fashion and finally approaches  a time-independent  spatially uniform profile   which is  consistent with a GGE state. 

In a similar way, the  exact Euler expression for the momentum density field $p(x,t)$ can be obtained. First we compute the momentum field $p^0(x',t)$ in the point particle picture and then transform to the momentum field for the hard rods using   
\begin{equation}
    p(x,t)=\frac{p^0(x',t)}{1+a\rho^0(x',t)}.
    \label{p^0-to-p}
\end{equation}
We find 
\begin{equation}
p^0(x',t)=\frac{4tT\rho_0}{(1-a\rho_0)\Big(1-2a\rho_0\Big)}\frac{1}{L}\sum_{k=1}^{\infty}\exp{-\frac{k^2\pi^2Tt^2}{2L^2(1-a\rho_0)^2}}\sin{\left(\frac{k\pi(1-2a\rho_0)}{2(1-a\rho_0)}\right)}\sin{\left(\frac{k\pi x'}{L(1-a\rho_0)}\right)},
\end{equation}
where $T=1$. In Fig.~\ref{fig:6}(b) we compare this with the results obtained from the MD simulations and we again see very good agreement. Here We observe that initially the momentum profile was zero everywhere. Once the gas is released, the rods with positive velocities near the middle of the box start moving to the right half. Thus the gas creates a positive momentum profile near the centre of the box. As time progresses, more  particles move to the right half and consequently the momentum profile spreads on both halves of the box. After some time,  of the order $L/\sqrt{T}$, finite size effects start showing, and some rods get reflected from the right walls. As a result, the motion of these rods start reducing the   momentum field. At very late times, each rod has undergone several collisions with both the walls and the gas equilibrates. The time scale of equilibration is also of the order $L/\sqrt{T}$. At this stage,  one has rods of opposite velocities with equal probabilities at any point of the box which leads  again to a zero momentum profile everywhere. 

Note that shifting the origin to $L/2-Na$ (i.e., $x' \rightarrow z'=x' -L/2+Na$) in the point particle problem, and taking $L,N \rightarrow \infty$ with $N/L=\rho_0$, one obtains the solution
of Euler GHD on the infinite line for times $t<<\frac{L}{\sqrt{T}}$ as:
\begin{align}
\begin{split}
\rho^0(z',t)&=\frac{2\rho_0}{2(1-2a\rho_0)}\text{erfc}\Big(\frac{z'}{\sqrt{2T}t}\Big), \\
p^0(z',t) &= \frac{2\rho_0\sqrt{T}}{(1-2a\rho_0)\sqrt{2\pi}}\text{exp}\Big(-\frac{z'^2}{2t^2T}\Big) ,
\end{split}
\end{align}
where $T=1$ and $\frac{2\rho_0}{1-2a\rho_0}$ is the initial density for $z'\in(-\infty,0)$. 
The early time plots (for $t<300$) in Fig.~\eqref{fig:6} can be obtained by transforming the above simpler functions to $\rho(x,t)$ and $p(x,t)$ using transformations in Eq.~\eqref{rho^0-to-rho}, \eqref{p^0-to-p} and Eq.~\eqref{map-x'-to-x}. The distortions of the densities of the point particles are appearing due to the non-linear transformations. 

\begin{figure}
\centering
\begin{subfigure}{.25\textwidth}
\centering
\includegraphics[width=1.0\linewidth]{{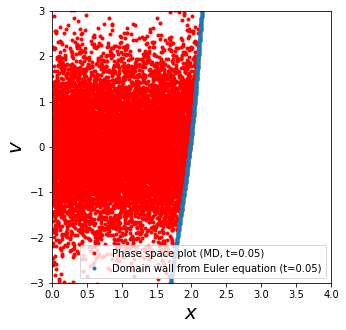}}
\caption{}
\end{subfigure}%
\begin{subfigure}{.25\textwidth}
\centering
\includegraphics[width=1.0\linewidth]{{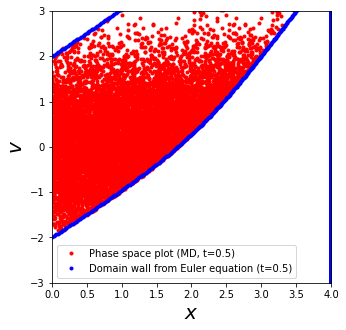}}
\caption{}
\end{subfigure}%
\begin{subfigure}{.25\textwidth}
\centering
\includegraphics[width=1.0\linewidth]{{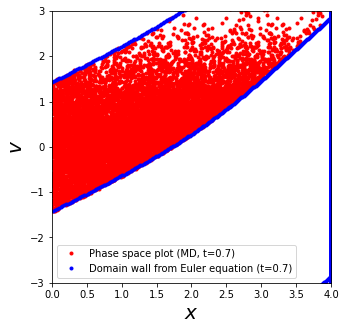}}
\caption{}
\end{subfigure}%
\begin{subfigure}{.25\textwidth}
\centering
\includegraphics[width=1.0\linewidth]{{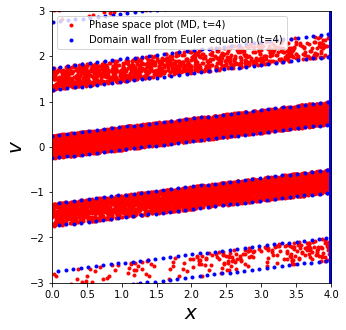}}
\caption{}
\end{subfigure}
\caption{{\bf Evolution of the domain lines for initial condition C}: Plot of the phase space distribution of the rods (red dots) for the free expansion problem at times (a) $t=0.05$ (b) $t=0.5$, (c) $t=0.7$ and (d) $t=4$. Solid blue lines represent the  domain lines obtained from the exact solution of the Euler equation. We see that the blue lines lie at the edges of the red region and are curved in contrast to the straight lines for the point particle [see \cite{chakraborti}]. At late times, the distribution wraps around the allowed region multiple times and thus creates fine structures. Here, $N=10000$, $L=4$, $a=0.0001$ \label{fig:9}}
\end{figure}

{\bf Domain line for initial condition C}: 
For the point particle case, $f^0(x',v,t)$ has a discontinuity in $x$ space (for a given $v$) for the free expansion problem. Since there is a mapping between the point particle Euler equation and the hard rod Euler equation, we expect that the Euler equation for hard rods will admit a  similar discontinuity. We call the line of discontinuity of $f(x,v,t)$ in the single-particle phase space as the "domain line". For the free expansion problem, the domain line can be found implicitly in the following manner. For times before the particles hit the right end of the container, the domain line for the point particle problem is given by $x'=vt+\frac{L}{2}-Na$. For general times (including times after the particles hit the right end of the container), we can do an analysis similar to \cite{chakraborti} to show that the single particle phase space distribution for the point particle problem for general times is given by:

\begin{equation}
    f^0(x',v,t)=\frac{2 \rho_0}{1-2a\rho_0}\frac{e^{-v^2/2T}}{\sqrt{2\pi T}}\sum_{n=-\infty}^{n=\infty}[\Theta(x'-vt-2n(L-Na)+L/2-Na)-\Theta(x'-vt+2n(L-Na)-L/2+Na)],
\end{equation}
where $n$ is an integer. From this, the domain line for the point particle problem can be computed as zeros of the argument of the theta functions appearing in the equation above. We know how $x'$ maps to $x$ ($x=x'+aF^0(x',t)$) within the framework of the Euler equation.
Thus we can compute the domain line for the hard rod problem as predicted by the Euler equation. We computed the domain line for the hard rod problem as predicted by the Euler equation, and plotted it (blue line) along with the phase space plot for the hard rods (red dots) predicted by the MD simulation. The plots are shown in Fig. \ref{fig:9}. We see that the blue line lies at the edge of the region occupied by the red dots. Thus the domain line predicted by Euler equation agrees with that predicted by MD simulation. We observe some key differences between hard rods (interacting integrable), alternate mass point-particle gas (non-integrable) and equal mass point particle gas (non-interacting integrable). In the hard rod gas, we see a sharp domain line which is not a straight line at early times (Fig.~\ref{fig:9}). In equal mass point particle gas discussed in  ~\cite{chakraborti}, a sharp and straight domain line was observed. However, in the alternate mass point particle gas, no sharp domain line was observed ~\cite{chakrabortiinteracting}.

\section{Conclusion}
\label{sec:dis}
In this paper we studied the macroscopic evolution of a collection of hard-rods in one dimension starting from three 
different initial conditions: (A) A uniformly filled box with an inhomogeneous velocity distribution --- half of the box is in thermal equilibrium and the other half has particles with a fixed velocity $v=1$, (B) One special particle with fixed velocity $v_0$ at the origin in the presence of a spatially uniform background of other  particles having thermal velocity distribution and (C) Free expansion from half of the box filled uniformly with thermal velocity distribution. For initial conditions (A) and (C) we find that the molecular dynamics results agree very well with the solutions of the Euler equations. However, for (A) we observe shocks at all times and find discrepancies from the Euler solutions at the location of the shocks which can be attributed  to the Navier-Stokes corrections to the Euler equations. For initial condition (B), the effect of the Navier-Stokes terms is more dramatic and here we show that the effect can be understood from the analytic solution of the Navier-Stokes equation. 

Our second important finding is the absence of GGE for initial conditions (A) and (B), whereas for initial condition (C) the system at late time approaches to a GGE state. The absence of GGE in the initial conditions (A) and (B) are manifested by the fact the density profile remains time dependent at all times. 

We find that the effect of the Navier-Stokes terms is very weak and to observe its effect one requires to have singular velocity distributions in the initial conditions such that a shock in the density profile survives for macroscopic time scale. At the location of the shock the large density gradient makes the contribution from the Navier-Stokes terms significant and consequently the solution near the shock (in Euler solution) becomes different from the Euler solutions. This is also what is observed in non-integrable systems \cite{joy,santhosh,sahil}. 

\par
Since the effect of dissipation is most noticeable near a shock, it is worth asking the question that for which initial conditions are shocks formed. It is easy to see there will be a shock only if the mapped point particle problem has a shock. This can be seen in the following way. Let $\delta x_0$ be the length scale over which the density is varying in the point particle problem. Then, using Percus' microscopic mapping, $\delta x=\delta x_0+a\delta N$, where $\delta x$ is the corresponding length scale in the hard rod problem, and $\delta N$ is the number of point particles in the length scale $\delta x_0$. If the point particle problem has a shock, then $\delta x_0$ will be small and $\delta N \sim O(1)$. Thus $\delta x$ will be of the order of few rod lengths, and there will be a shock in the hard rod problem also. If there is no shock in the point particle problem, then both $\delta x_0$ and $\delta N$ will be large, and hence $\delta x$ will also be large. Thus there will not then be any shock in the hard rod problem. We find that shocks in density profile of the point particles ( and hence of the hard rods) persists with time if  they start with singular velocity distributions [such as initial conditions (A) and (B)]. On the other hand if the rods have smooth velocity distribution to start with, then even if there are discontinuities in the density profile initially, the profiles at later times becomes smooth [such as initial condition (C)].

\par
For initial condition A, one may be curious what will happen if one chooses a Maxwellian distribution centered at $v=1$ instead of a $\delta$-function distribution [$h(v)=\delta(v-1)$] that we have choosen for the left half of the particles. If one does that, then the corresponding point particle problem will not have any shocks as both the halves will perform free expansion independently and we do not observe shocks in free expansion. If the spread of the Maxwellian is small enough, then one may observe a weak shock at small times, however the shock will weaken with time and eventually give rise to a smooth density profile. The time scale over which the shock will weaken will be inversely proportional to the spread of the Maxwellian, hence the shock for our initial condition A is infinitely long lived on the Euler level.

Observing the effect of the Navier-Stokes terms through the evolution of the density profile is difficult. However, it should get strongly  manifested in the evolution of Boltzmann's entropy \cite{chakraborti,chakrabortiinteracting,pandey2023boltzmann}, since the Euler solutions do not contribute to  entropy production. It would be interesting to study the entropy production for these initial conditions. However, identifying the contribution of the Navier-Stokes terms in the entropy production is still difficult because it involves taking appropriate combination of the limits of the coarse graining scale and the thermodynamic limit. This remains an interesting and challenging open problem.

\section{Acknowledgement}
AD and AK thank Benjamin Doyon for useful discussions.  AK would like to acknowledge the support of DST, Government of India Grant under Project No. ECR/2017/000634 and the MATRICS grant MTR/2021/000350 from the SERB, DST, Government of India. SS, AD and AK acknowledge the Department of Atomic Energy, Government of India, for their support under Project No. 19P1112\&D. HS and AD acknowledge the support from the
Science and Engineering Research Board (SERB, Government of India), under the VAJRA faculty scheme (No. VJR/2019/000079). 
AD and AK would like to thank the Isaac Newton Institute for Mathematical Sciences, Cambridge, for support and hospitality during the programme {\emph{New statistical physics in living matter: non equilibrium states under adaptive control}} where part of the work on this paper was undertaken.

\section{Appendix}
\appendix

\section{$\sqrt{t}$ behaviour of the shock for initial condition (A)}
\label{app:B}
In this appendix, we explain the $\sqrt{t}$ broadening of the shock found in Fig.~\ref{spreading-in-ICA}(a) for short times for initial condition A. For sufficiently short times, the evolution of the density profile  is  effectively the same as in an infinite box.
We thus consider a box $\Big[-\frac{L}{2},\frac{L}{2}\Big]$ with large $L$ and $N$ with $\rho_0=N/L$ fixed. The hard rods are arranged in initial condition A. This maps to a point particle problem in a box $\Big[-\frac{L}{2},\frac{L}{2}-Na\Big]$ with the left half having velocity $v=1$ and the right half having Maxwellian velocity distribution with unit temperature. Following the approach described in Sec.~\eqref{sec:3}, it can be shown that for short times:
\begin{equation}
    g'(x',t)=\frac{\rho_0}{1-a\rho_0}\Theta\Big(-x'-\frac{Na}{2}+v_0t\Big),
\end{equation}
\begin{equation}
    \rho_b'(x',t)=\frac{\rho_0}{2(1-a\rho_0)}\text{erfc}\Big[\frac{-x'-\frac{Na}{2}}{t\sqrt{2}}\Big].
\end{equation}
where $\rho_0=N/L$ and $v_0=1$. Following the same argument as in the Sec.~\ref{initialconditionB}, the width of the shock $\xi(t)$ is proportional to the fluctuation of number of background particles to the left of the shock. Thus:
\begin{equation}
    \xi^2(t)\propto\lim_{N,L\rightarrow\infty}\frac{a\rho_0}{2(1-a\rho_0)}\int_{-L/2}^{v_0t-Na/2}\text{erfc}\Big[\frac{-x'-\frac{Na}{2}}{t\sqrt{2}}\Big]dx'
\end{equation}
\begin{equation}
    =\lim_{N,L\rightarrow\infty}\frac{a\rho_0}{2(1-a\rho_0)}\int_{-L/2+Na/2}^{v_0t}\text{erfc}\Big[\frac{-u}{t\sqrt{2}}\Big]du
\end{equation}
\begin{equation}
    =\frac{a\rho_0}{2(1-a\rho_0)}\int_{-\infty}^{v_0t}\text{erfc}\Big[\frac{-u}{t\sqrt{2}}\Big]du
\end{equation}
\begin{equation}
    =\frac{a\rho_0t}{2(1-a\rho_0)}\int_{-v_0}^{\infty}\text{erfc}\Big[\frac{y}{\sqrt{2}}\Big]dy.
\end{equation}
where in going from second last line to last line, we have made a change of variable $y=-u/t$. The integral term in the last line is just some constant number. Thus $\xi(t)\propto\sqrt{t}$.

\bibliographystyle{unsrt}
\bibliography{references}

\begin{thebibliography}{10}

\bibitem{PhysRevLett.98.050405}
Marcos Rigol, Vanja Dunjko, Vladimir Yurovsky, and Maxim Olshanii.
\newblock Relaxation in a completely integrable many-body quantum system: An ab initio study of the dynamics of the highly excited states of 1d lattice hard-core bosons.
\newblock {\em Phys. Rev. Lett.}, 98:050405, Feb 2007.

\bibitem{Pozsgay_2013}
Balázs Pozsgay.
\newblock The generalized gibbs ensemble for heisenberg spin chains.
\newblock {\em Journal of Statistical Mechanics: Theory and Experiment}, 2013(07):P07003, jul 2013.

\bibitem{GGE}
Tim Langen, Sebastian Erne, Remi Geiger, Bernhard Rauer, Thomas Schweigler, Maximilian Kuhnert, Wolfgang Rohringer, Igor~E. Mazets, Thomas Gasenzer, and Jörg Schmiedmayer.
\newblock Experimental observation of a generalized gibbs ensemble.
\newblock {\em Science}, 348(6231):207--211, 2015.

\bibitem{Vidmar_2016}
Lev Vidmar and Marcos Rigol.
\newblock Generalized gibbs ensemble in integrable lattice models.
\newblock {\em Journal of Statistical Mechanics: Theory and Experiment}, 2016(6):064007, jun 2016.

\bibitem{cradle}
Trevor~Wenger Toshiya~Kinoshita and David~S. Weiss .
\newblock A quantum newton's cradle.
\newblock {\em Nature}, 2006(440):900--903, April 2006.

\bibitem{doi:10.1126/science.abf0147}
Neel Malvania, Yicheng Zhang, Yuan Le, Jerome Dubail, Marcos Rigol, and David~S. Weiss.
\newblock Generalized hydrodynamics in strongly interacting 1d bose gases.
\newblock {\em Science}, 373(6559):1129--1133, 2021.

\bibitem{doyonyoshimura}
Olalla~A. Castro-Alvaredo, Benjamin Doyon, and Takato Yoshimura.
\newblock Emergent hydrodynamics in integrable quantum systems out of equilibrium.
\newblock {\em Phys. Rev. X}, 6:041065, Dec 2016.

\bibitem{PhysRevLett.117.207201}
Bruno Bertini, Mario Collura, Jacopo De~Nardis, and Maurizio Fagotti.
\newblock Transport in out-of-equilibrium $xxz$ chains: Exact profiles of charges and currents.
\newblock {\em Phys. Rev. Lett.}, 117:207201, Nov 2016.

\bibitem{Alba_2021}
Vincenzo Alba, Bruno Bertini, Maurizio Fagotti, Lorenzo Piroli, and Paola Ruggiero.
\newblock Generalized-hydrodynamic approach to inhomogeneous quenches: correlations, entanglement and quantum effects.
\newblock {\em Journal of Statistical Mechanics: Theory and Experiment}, 2021(11):114004, nov 2021.

\bibitem{doyon1}
Benjamin Doyon.
\newblock {Lecture notes on Generalised Hydrodynamics}.
\newblock {\em SciPost Phys. Lect. Notes}, page~18, 2020.

\bibitem{doyon2}
Benjamin Doyon.
\newblock Generalized hydrodynamics of the classical toda system.
\newblock {\em Journal of Mathematical Physics}, 60(7):073302, 2019.

\bibitem{Bastianello_2021}
Alvise Bastianello, Andrea~De Luca, and Romain Vasseur.
\newblock Hydrodynamics of weak integrability breaking.
\newblock {\em Journal of Statistical Mechanics: Theory and Experiment}, 2021(11):114003, nov 2021.

\bibitem{Bastianello_2022}
Alvise Bastianello, Bruno Bertini, Benjamin Doyon, and Romain Vasseur.
\newblock Introduction to the special issue on emergent hydrodynamics in integrable many-body systems.
\newblock {\em Journal of Statistical Mechanics: Theory and Experiment}, 2022(1):014001, jan 2022.

\bibitem{10.1063/1.5018624}
Herbert Spohn.
\newblock {Interacting and noninteracting integrable systems}.
\newblock {\em Journal of Mathematical Physics}, 59(9), 06 2018.
\newblock 091402.

\bibitem{LPS}
J.~L. Lebowitz, J.~K. Percus, and J.~Sykes.
\newblock Time evolution of the total distribution function of a one-dimensional system of hard rods.
\newblock {\em Phys. Rev.}, 171:224--235, Jul 1968.

\bibitem{spohnbook}
Herbert Spohn.
\newblock {\em {\it Large Scale Dynamics of Interacting Particles}}.
\newblock Springer Berlin, Heidelberg, December 2012.

\bibitem{boldighrini2}
C.~Boldrighini and Y.M. Suhov .
\newblock One-dimensional hard-rod caricature of hydrodynamics: “navier–stokes correction” for local equilibrium initial states.
\newblock {\em Communications in Mathematical Physics}, 1997(189):577--590, November 1997.

\bibitem{doyonspohn}
Benjamin Doyon and Herbert Spohn.
\newblock Dynamics of hard rods with initial domain wall state.
\newblock {\em Journal of Statistical Mechanics: Theory and Experiment}, 2017(7):073210, jul 2017.

\bibitem{ferrari2023}
Pablo~A Ferrari and Stefano Olla.
\newblock Macroscopic diffusive fluctuations for generalized hard rods dynamics.
\newblock {\em arXiv preprint arXiv:2305.13037}, 2023.

\bibitem{PhysRevLett.106.140405}
Amy~C. Cassidy, Charles~W. Clark, and Marcos Rigol.
\newblock Generalized thermalization in an integrable lattice system.
\newblock {\em Phys. Rev. Lett.}, 106:140405, Apr 2011.

\bibitem{PhysRevLett.106.227203}
Pasquale Calabrese, Fabian H.~L. Essler, and Maurizio Fagotti.
\newblock Quantum quench in the transverse-field ising chain.
\newblock {\em Phys. Rev. Lett.}, 106:227203, Jun 2011.

\bibitem{PhysRevLett.127.020501}
J.~Eisert.
\newblock Entangling power and quantum circuit complexity.
\newblock {\em Phys. Rev. Lett.}, 127:020501, Jul 2021.

\bibitem{De2018}
Jacopo De~Nardis, Denis Bernard, and Benjamin Doyon.
\newblock Hydrodynamic diffusion in integrable systems.
\newblock {\em Physical review letters}, 121(16):160603, 2018.

\bibitem{Percus}
J.~K. Percus.
\newblock Exact solution of kinetics of a model classical fluid.
\newblock {\em The Physics of Fluids}, 12(8):1560--1563, 1969.

\bibitem{BernsteinPercus}
M.~Bernstein and J.~K. Percus.
\newblock Expansion into a vacuum: A one-dimensional model.
\newblock {\em Phys. Rev. A}, 37:1642--1653, Mar 1988.

\bibitem{jepsen1965dynamics}
D.~W. Jepsen.
\newblock Dynamics of a simple many-body system of hard rods.
\newblock {\em Journal of Mathematical Physics}, 6(3):405--413, 1965.

\bibitem{chakraborti}
Subhadip Chakraborti, Abhishek Dhar, Sheldon Goldstein, Anupam Kundu, and Joel~L Lebowitz.
\newblock Entropy growth during free expansion of an ideal gas.
\newblock {\em Journal of Physics A: Mathematical and Theoretical}, 55(39):394002, sep 2022.

\bibitem{chakrabortiinteracting}
Subhadip Chakraborti, Abhishek Dhar, and Anupam Kundu.
\newblock {Boltzmann’s Entropy During Free Expansion of an Interacting Gas}.
\newblock {\em Journal of Statistical Physics}, 190(74), 03 2023.

\bibitem{pandey2023boltzmann}
Saurav Pandey, Junaid~Majeed Bhat, Abhishek Dhar, Sheldon Goldstein, David~A. Huse, Manas Kulkarni, Anupam Kundu, and Joel~L. Lebowitz.
\newblock Boltzmann entropy of a freely expanding quantum ideal gas, 2023.

\bibitem{joy}
Jilmy~P. Joy, Sudhir~N. Pathak, and R.~Rajesh.
\newblock Shock propagation following an intense explosion: Comparison between hydrodynamics and simulations.
\newblock {\em Journal of Statistical Physics}, 182, Feb 2021.

\bibitem{santhosh}
Subhadip Chakraborti, Santhosh Ganapa, P.~L. Krapivsky, and Abhishek Dhar.
\newblock Blast in a one-dimensional cold gas: From newtonian dynamics to hydrodynamics.
\newblock {\em Phys. Rev. Lett.}, 126:244503, Jun 2021.

\bibitem{sahil}
Sahil~Kumar Singh, Subhadip Chakraborti, Abhishek Dhar, and P.~L. Krapivsky.
\newblock Blast waves in the zero temperature hard sphere gas: Double scaling structure.
\newblock {\em Journal of Statistical Physics}, 190, Jul 2023.

\end{thebibliography}



\end{document}